\documentstyle{article}
\begin{document}

\begin{center}
\baselineskip=1\baselineskip
~\\
{\Large {\bf SELF-SIMILAR EQUILIBRIA OF SELF-GRAVITATING, MAGNETIZED, ROTATING, ISOTHERMAL SYSTEMS\footnote{To appear in ApJ: v556 n1 Jul 20, 2001}}}
~\\
~\\
~\\
{\sc { MOHSEN SHADMEHRI\footnote{shad@science1.um.ac.ir}\\   AND   \\JAMSHID GHANBARI\footnote{ghanbari@science1.um.ac.ir}}}\\
{\it Department of Physics, School of Sciences, Ferdowsi University, Mashhad, Iran}
~\\
~\\
~\\
Accepted by {\it The Astrophysical Journal}
~\\

\end{center}
~\\
\baselineskip=1\baselineskip
\begin{abstract}
\baselineskip=1\baselineskip
The self-similar equilibrium models of self-gravitating, rotating, isothermal systems are investigated analytically. In these models the rotation velocity is constant and the density varies as $\frac{f(\theta, \varphi)}{r^2}$, where $r$ and $\theta$ are the spherical radius and the co-latitude, respectively. The nonaxisymmetric solutions contain three free parameters, one of the parameters depends on the rotation velocity. These parameters determine the overall  shape of the density distribution. By assuming that the dominant component of the magnetic field is purely toroidal and the ratio of the purely toroidal magnetic pressure to the gas pressure, $\alpha$, is spatially constant, the axisymmetric solutions generalized so as the effect of magnetic field could be studied. We find that the equilibria of axially symmetric systems yield ellipsoids or spheres only when the ratio of rotation velocity to the sound speed is taken to be $\sqrt{2\alpha}$.
~\\
{\it Subject headings}: hydromagnetics - stars: kinematics - stars: formation - galaxies: structure
\end{abstract}

~\\
~\\

\noindent{\bf 1. INTRODUCTION}
~\\
Understanding the equilibrium structure of self-gravitating isothermal systems is a fundamental issue in astrophysics. In star-forming regions many structures have been identified from sheets, to filaments, to elongated clumps (Kulkarni \&\ Heiles 1988; Jijina, Myers, \&\ Adams 1999). By some analytical solutions, the hydrostatic equilibrium of these structures have been studied previously (Shu 1992; Ostriker 1964). Several authors, however, investigated the problem by numerical techniques so as to include the effects of rotation and magnetic fields (see, e.g., Fiege \&\ Pudritz 2000; Curry 2000; Tomisaka, Ikeuchi, \&\ Nakamura 1988). Many models of newborn stars built upon these equilibrium configurations (e.g., Kiguchi et al. 1987). Also, there is a close analogy between gaseous self-gravitating equilibrium objects and collisionless stellar systems (see, e.g., Binney \&\ Tremaine 1987). Thus the equilibria of isothermal systems can be used for determining the steady-state distribution function of collisionless stellar systems (Toomre 1982; Hayashi, Narita, \&\ Miyama 1982).

With these considerations, recently Medvedev and Narayan (2000; hereafter MN) classified and derived analytically self-similar axisymmetric equilibria of a self-gravitating isothermal system. They presented two families of equilibria: (1) Cylindrically symmetric solutions, in which all quantities depend on the cylindrical radius only. (2) Axially symmetric solutions, in which the quantities vary as functions of the spherical radius and the co-latitude. In fact, these axially symmetric solutions are very interesting. These equilibria form a two-parameter family of solutions. One of the parameters depends on the rotation velocity and the other one controls the symmetry of the solution with respect to the equatorial plane. Nevertheless, this study like most previous work has focused on the case of axisymmetric equilibria of self-gravitating systems.The absence in the literature of $\it nonaxisymmetric$ equilibrium solution suggests that a fresh theoretical approach is necessary.

Our motivation for this work is to determine nonaxisymmetric equilibria of self-gravitating systems. We extend the work of MN by obtaining nonaxisymmetric self-similar equilibria of a self-gravitating system analytically. Because of great complexity and intrinsically nonlinear nature of the problem, any analytical solution is unique and, thus,of great scientific value. Our nonaxisymmetric equilibrium configurations form a three-parameter family of solutions and in the case of axisymmetric equilibrium, these solutions reduce to the solutions of MN. Furthermore, we shall study the effect of magnetic fields on the axisymmetric self-similar equilibria solutions. Since we are interested in treating the problem analytically and for simplicity, we assume that the dominant component of the magnetic field is purely toroidal and the ratio of the magnetic pressure to the gas pressure, $\alpha$, is spatially constant. MN showed that the non-rotating axially symmetric solutions form confocal ellipsoids or spheres. But we find that the non-rotating, magnetized, axially symmetric solutions tend to the cylindrically symmetric structures. However, the equilibria of axially symmetric systems both with rotation and toroidal magnetic field yield ellipsoids or spheres only when the ratio of rotation velocity to the sound speed is taken to be $\sqrt{2\alpha}$.

In $\S 2$, we present the general formulation and basic assumptions. We derive a three-parameter family of nonaxisymmetric self-similar solutions in $\S 3$ and investigate the properties of these solutions in $\S 4$. We summarize our results in $\S 5$.
~\\
~\\
\noindent{\bf 2. GENERAL FORMULATION}\\
We present here the basic equations used to describe the equilibrium structure of a rotating self-gravitating system permeated by a toroidal magnetic field. We assume that the medium is an isothermal ideal gas and the magnetic field is frozen into the gas. However, the assumption that the system is isothermal is good for low-density region. In spherical coordinates $(r, \theta, \varphi)$, the basic equations are magnetohydrostatic equation, the Poisson equation and the equation of state:
\begin{equation}
\frac{1}{\rho}\frac{\partial p}{\partial r}+\frac{\partial \Psi}{\partial r}+\frac{1}{4\pi r}\frac{B_{\varphi}}{\rho}\frac{\partial}{\partial r}(r B_{\varphi})=\frac{v_{\varphi}^{2}}{r},
\end{equation}
\begin{equation}
\frac{1}{\rho}\frac{\partial p}{\partial \theta}+\frac{\partial \Psi}{\partial \theta}+\frac{1}{4\pi  sin \theta}\frac{B_{\varphi}}{\rho}\frac{\partial}{\partial \theta}(B_{\varphi} sin \theta)=v_{\varphi}^{2} cot \theta,
\end{equation}
\begin{equation}
\frac{1}{\rho}\frac{\partial p}{\partial \varphi}+\frac{\partial \Psi}{\partial \varphi}=-v_{\varphi}\frac{\partial v_{\varphi}}{\partial \varphi},
\end{equation}
\begin{equation}
\frac{1}{r^{2}}\frac{\partial}{\partial r}(r^{2}\frac{\partial\Psi}{\partial r})+\frac{1}{r^{2} sin \theta}\frac{\partial}{\partial \theta}(sin \theta \frac{\partial \Psi}{\partial \theta})+\frac{1}{r^{2} sin^{2}\theta}\frac{\partial^{2}\Psi}{\partial\varphi^{2}}=4\pi G\rho,
\end{equation}
\begin{equation}
\\p=c_{s}^{2}\rho,
\end{equation}
where $\rho$, $p$, $v_{\varphi}$, $c_{s}$, $B_{\varphi}$ and $\Psi$ denote the gas density, pressure,toroidal component of the velocity due to the rotation, sound speed, toroidal magnetic field, and gravitational potential, respectively. We assumed that the entire system is rotating around a common axis which specifies the axis of the spherical coordinate system. It is clear that if $\Omega$ denotes the angular velocity, then $v_{\varphi}=\Omega r sin \theta$.  We note that purely toroidal magnetic field under the cylindrical symmetry, $B_{\varphi}(r, \theta)$, automatically satisfies $\nabla.\bf \it B=0$. In fact this configuration of the magnetic field requires a rather artificial current configuration. Namely the current must flow along the rotation axis (z-axis). Although systems with such fields are not too common in astrophysics, our results provide an intuitive way to understand more complex systems. Also, the rotation velocity, the density and gravitational potential, in general, are functions of $r$, $\theta$, and $\varphi$.

Solving equations (1) to (5), in general, is a difficult work. Thus to simplify the problem, we need another constraint. As for molecular clouds, observations suggest that the magnetic field strength often varies with the density roughly according to $B \sim \rho^{\frac{1}{2}}$ (e.g., Heiles et al. 1993). It implies that the ratio of magnetic pressure to the thermal pressure, $\alpha$, is spatially constant:
\begin{equation}
\alpha=\frac{B_{\varphi}^{2}}{8\pi p},
\end{equation}
where in molecular clouds observations show that $0<\alpha<10$ (Heiles et al. 1993). For our problem we use it as a free parameter. By changing the value of $\alpha$, the effect of purely toroidal magnetic field on the equilibrium state of self-gravitating systems can be studied. It must be noted that by using this constraint we can investigate effects of magnetic field $\it only$ on the $\it axisymmetric$ solutions. In fact our nonaxisymmetric solutions are nonmagnetized.

Now, we introduce the dimensionless variables according to
\begin{equation}
\rho\rightarrow\hat{\rho}\rho, r\rightarrow\hat{r} r, \Psi\rightarrow\hat{\Psi}\Psi,v_{\varphi}\rightarrow\hat{v}v_{\varphi}, B_{\varphi}\rightarrow\hat{B} B_{\varphi},
\end{equation}
where
\begin{equation}
\hat{\rho}=\rho_{0}, \hat{r}=(4\pi G\rho_{0})^{-\frac{1}{2}}c_{s}^{2}, \hat{\Psi}=c_{s}^{2}, \hat{v}=c_{s}, \hat{B}=(4\pi\rho_{0})^{\frac{1}{2}} c_{s}.
\end{equation}
By transforming to dimensionless variables and using equation (6), equations (1) to (4) are rewritten as the following:
\begin{equation}
\frac{\partial\Psi}{\partial r}=\frac{v_{\varphi}^{2}-2\alpha}{r}-(\alpha+1)\frac{\partial ln \rho}{\partial r},
\end{equation}
\begin{equation}
\frac{\partial\Psi}{\partial\theta}=(v_{\varphi}^{2}-2\alpha) cot \theta -(\alpha+1)\frac{\partial ln \rho}{\partial \theta},
\end{equation}
\begin{equation}
\frac{\partial\Psi}{\partial\varphi}= - \frac{\partial ln \rho}{\partial \varphi}- \frac{1}{2}\frac{\partial}{\partial\varphi}(v_{\varphi}^{2}),
\end{equation}
\begin{equation}
\frac{1}{r^{2}}\frac{\partial}{\partial r}(r^{2}\frac{\partial\Psi}{\partial r})+\frac{1}{r^{2} sin \theta}\frac{\partial}{\partial \theta}(sin \theta \frac{\partial \Psi}{\partial \theta})+\frac{1}{r^{2} sin^{2}\theta}\frac{\partial^{2}\Psi}{\partial\varphi^{2}}=\rho,
\end{equation}
By substituting equations (9), (10), and (11) into equation (12) we obtain
\begin{equation}
\frac{\alpha+1}{r^{2}}\frac{\partial}{\partial r}(r^{2}\frac{\partial ln\rho}{\partial r})+\frac{\alpha+1}{r^{2}sin \theta}\frac{\partial}{\partial \theta}(sin \theta \frac{\partial ln \rho}{\partial \theta})+\frac{1}{r^{2}sin^{2}\theta}\frac{\partial^{2}ln \rho}{\partial \varphi^{2}}+\rho=K(r, \theta, \varphi),
\end{equation}
\begin{equation}
\\K(r, \theta, \varphi)=\frac{1}{r^{2}}\frac{\partial}{\partial r}(r v_{\varphi}^{2})+\frac{1}{r^{2}sin \theta}\frac{\partial}{\partial \theta}(v_{\varphi}^{2} cos\theta)-\frac{1}{r^2 sin^2\theta}\frac{\partial^2}{\partial\varphi^2}(v_{\varphi}^2).
\end{equation}
It would be difficult to study systematically the very large space of all possible solutions of these equations. In fact, we need some information about the density, $\rho$, and the rotation velocity, $v_{\varphi}$. For example, by knowing the form of one of the variables it is possible to find another variable. But here we are interested in self-similar solutions which will be studied in the next section. These solutions are interpreted with much less effort than would be involved in a direct attempt to solve the full set of partial differential equations.
~\\
~\\
\noindent{\bf 3. SELF-SIMILAR SOLUTIONS}\\
We introduce the self-similar forms for the density and the rotation velocity as 
\begin{equation}
\rho(r, \theta, \varphi)=\frac{f(\theta, \varphi)}{r^{\nu_{1}}}, v_{\varphi}=\frac{v_{0 \varphi}}{(r sin\theta)^{\nu_{2}}},
\end{equation}
where $\nu_{1}$, $\nu_{2}$, and $f(\theta, \varphi)$ to be calculated. The same self-similar solutions for the axisymmetric case has been introduced by MN. Their angular part of the density distribution was a function of $\theta$ only. Since we are going to study the nonaxisymmetric solutions, the angular part of the density is a function of both $\theta$ and $\varphi$. Substituting the self-similar forms into equations (13) and (14), we obtain
\begin{equation}
\nu_{1}=2, \nu_{2}=0,
\end{equation}
and $f(\theta, \varphi)$ can be found by solving this equation:
\begin{equation}
\\-2(\alpha+1)+\frac{\alpha+1}{sin \theta}\frac{\partial}{\partial \theta}(sin \theta \frac{\partial ln f}{\partial \theta})+\frac{1}{sin^{2}\theta}\frac{\partial^{2}ln f}{\partial \varphi^{2}}+f=0.
\end{equation}

This is the main equation which must be solved. We note that for the nonmagnetized and axisymmetric case, this equation reduces to the equation which has been solved by MN. To solve equation (17), we use transformations which are equivalent to Howard's transformation used by Toomre (1982) and MN. First, we rewrite this equation as follows,
\begin{equation}
\frac{\alpha+1}{sin \theta}\frac{\partial}{\partial \theta}[sin \theta \frac{\partial}{\partial \theta}ln(f sin^{2}\theta)]+\frac{1}{sin^{2}\theta}\frac{\partial^{2}ln f}{\partial \varphi^{2}}+f=0.
\end{equation}
Upon introducing a new function, $w$, and a new independent variable, $\xi$, as follows,
\begin{equation}
\\w=ln(f sin^{2}\theta), \xi=ln\vert tan(\frac{\theta}{2})\vert^{\frac{1}{\sqrt{\alpha+1}}},
\end{equation} 
equation (18) is greatly simplified and becomes
\begin{equation}
\frac{\partial^{2}w}{\partial\xi^{2}}+\frac{\partial^{2}w}{\partial\varphi^{2}}+e^{w}=0.
\end{equation}

This equation is similar to the Lane-Emden equation which has been written in cartesian coordinates. Schmid-Burgk (1967) and independently Stuart (1967) presented an interesting two-dimensional solution of the Lane-Emden equation in cartesian coordinates. Thus, in analogy to their solution we can write an analytical solution for equation(20),
\begin{equation}
\\w=ln\lbrace\frac{2a^{2}(1-C^{2})}{[cosh(a\xi+b)+C cos(a\varphi)]^{2}}\rbrace,
\end{equation}
where $a$, $b$, and $C$ are free parameters. Recalling the transformations (19), we finally obtain an analytical solution,
\begin{equation}
\\f(\theta, \varphi)=\frac{1}{sin^{2}\theta}\frac{2a^{2}(1-C^{2})}{[cosh(ln  \vert tan\frac{\theta}{2}\vert^{\frac{a}{\sqrt{\alpha+1}}}+b)+C cos(a\varphi)]^{2}}.
\end{equation}
This solution can be simplified further and the densiy becomes
\begin{equation}
\rho(r, \theta, \varphi)=\frac{2A^{2}B(1-C^{2})}{r^{2}sin^{2}\theta}\frac{\vert tan\frac{\theta}{2}\vert^{\frac{A}{\sqrt{\alpha+1}}}}{(1+B\vert tan\frac{\theta}{2}\vert^{\frac{A}{\sqrt{\alpha+1}}}+2C\sqrt{B}\vert tan\frac{\theta}{2}\vert^{\frac{A}{2\sqrt{\alpha+1}}}cos\frac{A}{2}\varphi)^{2}},
\end{equation}
where $A=2a$, $B=e^{2b}$, and $-1<C<1$. This solution describes the nonaxisymmetric density distribution of a self-gravitating system. It is clear that to obtain the nonmagnetized axisymmetric equilibria we must set $\alpha=0$ and $C=0$ and our solution reduces to the solution of MN. Equation (23) presents a three-parameter family of solutions (note that for nonaxisymmetric we must set $\alpha=0$) and is smooth and well-behaved for all values of $\theta$ and $\varphi$ in the domains $0<\theta<\pi$ and $0\leq\varphi\leq2\pi$. However, the equation is ill-defined at $r=0$ or $\theta=0$. We can investigate if a singular solution exists at the origin or on the axis. Using Gauss'theorem, MN calculated these additional mass densities for the case of the nonmagnetized axisymmetric equilibria. The reader is referred to MN for a full derivation. Following MN, we find that our solution has no singular mass at the origin. But the requirement of a non-singular density on the axis gives,
\begin{equation}
\\A=\frac{2+v_{0\varphi}^{2}}{\sqrt{\alpha+1}}.
\end{equation}
This equation shows that the rotation velocity and the magnetic field determine the value of A. Thus, the solution (23) becomes
\begin{equation}
\rho(r, \theta, \varphi)=\frac{2A^{2}B(1-C^{2})}{r^{2}sin^{2}\theta}\frac{\vert tan\frac{\theta}{2}\vert^{A_{ef}}}{(1+B\vert tan\frac{\theta}{2}\vert^{A_{ef}}+2C\sqrt{B}\vert tan\frac{\theta}{2}\vert^{\frac{A_{ef}}{2}}cos\frac{A}{2}\varphi)^{2}},
\end{equation}
where
\begin{equation}
\\A_{ef}=\frac{2+v_{0\varphi}^{2}}{\alpha+1}.
\end{equation}
Since the exponent of $tan\frac{\theta}{2}$ determines the overall shape of the density distribution, equation (26) is very important. In the next section we study the properties of the solutions.
~\\
~\\
\noindent{\bf 4. PROPERTIES OF THE SOLUTIONS}\\
Since a real self-gravitating system is finite, we truncated the system by an external pressure. The external medium is assumed to be non-self-gravitating and of negligible density. In this way we can define the surface of the system. If $p_{s}$ represent the external pressure in a nondimentional form, we obtain the equation of the surface
\begin{equation}
\\r_{s}(\theta, \varphi)=\sqrt{\frac{2A^{2}B(1-C^{2})}{p_{s}}}\frac{1}{sin\theta}\frac{\vert tan\frac{\theta}{2}\vert^{\frac{A_{ef}}{2}}}{1+B\vert tan\frac{\theta}{2}\vert^{A_{ef}}+2C\sqrt{B}\vert tan\frac{\theta}{2}\vert^{\frac{A_{ef}}{2}}cos\frac{A}{2}\varphi}.
\end{equation}
Notice also that for $\alpha=0$ and $C=0$, this equation reduces to the solution which MN obtained. As equations (24) and (26) show, the rotation velocity, $v_{0\varphi}$, and the ratio of magnetic pressure to the gas pressure, $\alpha$, determine the values of $A$ and $A_{ef}$. But it is possible to make another constraint: Since we are interested in the solutions which have no discontinuity on the surface, we accept those values of $A$ that $r_{s}(\theta, \varphi=0)=r_{s}(\theta, \varphi=2\pi)$. Thus,
\begin{equation}
\\A=2\nu,
\end{equation}
where $\nu$ is a positive integer. So only a limited number of values of $v_{0\varphi}$ and $\alpha$ gives acceptable values of $A$.

As mentioned before, the value of $A_{ef}$ determines the overal shape of the system. For the nonmagnetized case, we have $A_{ef}=A$. So, only the rotation velocity determines the value of $A$. As the $v_{0\varphi}$ increases, the value of $A$ becomes larger. Equation(26) shows that as $\alpha$ increases, the value of $A_{ef}$ decreases. It seems that a purely toroidal magnetic field with a constant ratio of the magnetic pressure to the thermal pressure essentially cancels the effects of rotation. With these considerations, we focus this current discussion on $\it nonmagnetized$ and $\it magnetized$ solutions.
~\\
~\\
\noindent{\it4.1. Nonmagnetized Equilibria}\\
First, we study the nonrotating solutions: $A_{ef}=A=2$. The solution (27) reads
\begin{equation}
\\r_{s}(\theta, \varphi)=\frac{\lambda}{1+\epsilon cos \theta+\delta sin \theta cos \varphi},
\end{equation}
where
\begin{equation}
\lambda=\sqrt{\frac{8(1-C^{2})}{p_{s}(1+B)^{2}}}, \epsilon=\frac{1-B}{1+B}, \delta=\frac{2C\sqrt{B}}{1+B}.
\end{equation}
Note that the case $C=0$ and $B\neq 1$  corresponds to a prolate ellipsoid and if $B=1$ this ellipsoid reduces to a singular sphere. These axisymmetric equilibria have been studied extensively by MN. They showed that the solutions with $B>1$ determine prolate configurations which are shifted upwards along the $z$ axis and those with $B<1$ are shifted downwards. So, the parameter $B$ controls the symmetry of the solutions with respect to the equatorial plane.

It can easily be verified that the general case $C\neq0$ and $B\neq 1$ represents a family of ellipsoids. By rotating the coordinate system $(x, y, z)$ about $y$ axis from the positive direction of $z$ axis to the positive direction of $x$ axis we can define a new coordinate system $(X, y, Z)$. Assuming that $\omega$ is the angle of rotation, upon straightforward but cumbersome algebric manipulations, we can rewrite equation (29) in a new coordinate system $(X, y, Z)$ 
\begin{equation}
\frac{(X-\frac{\lambda \beta}{1-\beta^{2}})^{2}}{\frac{\Gamma^{2}}{1-\beta^{2}}}+\frac{y^{2}}{\Gamma^{2}}+\frac{(Z+\frac{\lambda \gamma}{1-\gamma^{2}})^{2}}{\frac{\Gamma^{2}}{1-\gamma^{2}}}=1,
\end{equation} 
where
\begin{equation}
\beta=\epsilon sin\omega - \delta cos \omega, 
\gamma=\epsilon cos \omega + \delta sin \omega, 
\omega=\frac{1}{2}tan^{-1}\frac{2\epsilon\delta}{\epsilon^2-\delta^2}, 
\end{equation}
and
\begin{equation}
\Gamma^2=\frac{(1-\beta^2\gamma^2)\lambda^2}{(1-\beta^2)(1-\gamma^2)}.
\end{equation}

Equation (31) shows that the case $C=0$ and $B\neq 1$ represents a family of ellipsoids, with traces on the $y-z$ and $x-z$ planes which are ellipses and traces on the $x-y$ plane which are circles with centers at origin. Since $r_{s}(\theta)$ is not invariant under the transformation $\theta\rightarrow\pi-\theta$, this configuration is not symmetric with respect to the equatorial plane.On the other hand, the case $B=1$ and $C\neq0$ corresponds to a family of ellipsoids which are elongated along the $x$ axis and $r_{s}(\theta, \varphi)$ is invariant under the transformations $\theta\rightarrow\pi-\theta$ and $\varphi\rightarrow\varphi$. Thus, the equatorial plane is the symmetry plane and projections of these ellipsoids onto the $y-z$ plane are circles with centers at origin. In fact, the general solution (27) shows that $r_{s}(\theta, \varphi)$ is invariant under these transformations only for $B=1$. So, the parameter $B$ in our nonaxisymmetric solution controls the symmetry of the solutions with respect to the equatorial plane.

As mentioned before, we restrict our study only to the solutions which have no singular mass on the axis. This constraint gives $A\succeq 2$ or $\nu\succeq 1$; note that the case $\nu=1$ has already been studied. For $C=0$ and $A>2$, equation (27) corresponds to a family of toroidal structures in which the parameter $B$ determines the asymmetry of these equilibria with respect to the equatorial plane. In this case, as $A$ increases, the systems become flattend and tend to a disk. It seems that in our solution the parameter $C$ alone determines the asymmetry of the system with respect to the $z$ axis. Although it is true, the parameter $A$ (or $\nu$) is also very important. We see that the equation (27) is invariant under the transformations $\theta\rightarrow\theta$ and $\varphi\rightarrow\pi+\varphi$ only for even values of $\nu$. Therefore, in the case $C\neq 0$ we may have configurations which are symmetric with respect to the $z$ axis. The rotation velocity has a fundamental role in determining the overal shape of the density distribution. For small values of $C$ and $A>2$, the toroidal equilibrium structures are elongated in some directions. We refer to these elongated parts to as $\it knobs$. As discussed above, the number of these knobs depends on the rotation velocity. For example, the case $A=8$ $(\nu=4)$ represent a toroidal configuration with four knobs. With increasing $C$ and tending the parameter to the unity, these knobs become larger and the toroidal part becomes smaller.
~\\
~\\
\noindent{\it4.2. Magnetized Equilibria}\\
It is useful to study the effect of the magnetic field on our solutions. We postulated that the magnetic field structure is purely toroidal and corresponds to that of constant ratio of the magnetic pressure to the thermal pressure. Although purely toroidal fields are probably far from reality, this is the best way to investigate the effects of the magnetic fields on the equilibria analytically.

Now, we consider the effects of the magnetic field on axisymmetric solutions: $C=0$. Equation (26) shows that a nonrotating system always has $A_{ef}<2$ and the value of $A_{ef}$ decreases with increasing the strength of magnetic field. While the nonrotating and nonmagnetized  systems are ellipsoids, the nonrotating and magnetized systems tend to the cylindrically symmetric configurations. Also, if we consider rotation such that $v_{0\varphi}<\sqrt{2\alpha}$, the system has cylindrically symmetric structure.

If $v_{0\varphi}=\sqrt{2\alpha}$, we have $A_{ef}=2$ and the equilibria are like those which have been investigated in the previous subsection for the case $C=0$. Finally, the case $v_{0\varphi}>\sqrt{2\alpha}$ and $C=0$ corresponds to a family of toroidal configurations. We know that as $v_{0\varphi}\rightarrow\infty$ the equilibria tend to a disk. However, the magnetic field decreases the effect of rotation and in this case it causes the system to become a toroidal structure not a disk.
~\\
~\\
\noindent{\bf 5. DISCUSSION AND SUMMARY} \\
Self-similar equilibria of a self-gravitating rotating system containing a purely toroidal magnetic field has been investigated. A three-parameter family of solutions for nonmagnetized self-gravitating systems have been found. By assuming that the ratio of the magnetic pressure to the gas pressure, $\alpha$, is spatially constant, the axisymmetric solutions were generalized so that the effect of the magnetic field could be studied. We have shown that, depending on the values of these parameters, the overall behavior of the density distribution changes.

Recently, Galli et al. (2001) studied binary and multiple star formation by considering the nonaxisymmetric equilibria of self-gravitating, magnetized, differentially-rotating, completely flattened singular isothermal disks with critical or supercritical ratios of mass-to-flux. They found that lopsided configurations exist at any dimensionless rotation rate. Also, multiple-lobed $(M = 2, 3, 4, ...)$ configurations correspond to rotations of the equilibrium configurations by multiples of $\frac{\pi}{M}$ and bifurcations into sequences with $M = 2, 3, 4, 5,$ and higher symmetry require considerable larger rotation rates. These results are in a well agreement with our analysis, at least qualitatively. We must note that the flattening of our nonmagnetized equilibria produced by rotation rather than by magnetic fields. 

As equation (25) shows the parameters $A$ or $A_{ef}$ determine the overall shape of density distribution. We showed that only a limited number of values of $v_{0\varphi}$ and $\alpha$ give acceptable values for $A$ or $A_{ef}$. Since for nonmagnetized solutions we have $A=A_{ef}$, as the value of  rotation velocity increases, the value of $A$ becomes larger. Equation (25) shows that for $A>2$, the toroidal equilibrium structures have knobs and the number of these knobs depends on the rotation velocity: as rotation velocity increases, the number of the knobs increases. It is interesting that in the case $C\neq 0$, some values of $A$ construct equilibrium configurations which are symmetric with respect to the $z$ axis. We see that our multiple-lobed $(\nu = 1, 2, 3, ...)$ equilibria require larger rotation rates. The equilibria solutions become flattened as the rotation velocity increases and they tend to a thin disk as $A \rightarrow \infty$. Thus, our nonaxisymmetric solution tend to a nonaxisymmetric thin disk as the rotation velocity increases. These configurations are equivalent to the nonaxisymmetric solutions which have been obtained by Galli et al. (2001). Detailed observations of the gas disks associated with bipolar outflows indicate that the disks are rotating very fast (Kaifu 1987). The rotation period is comparable to the free-fall time scale of the gas disks and it means that the disk is near the rotational balance. Thus, in these systems the effect of rotation should have been important in the formation of the central star. On the other hand, it is very important that the formation of multiple stellar systems could never result from any calculation that imposes a priori an assumption of axial symmetry. So, as long as the starting conditions are  the nonaxisymmetric disk solutions, gravitational collapses seem in general to produce fragmentation. 

For $B=1$ and $C=0$, the solutions have been obtained by Toomre (1982) and Hayashi, Narita, \&\ Miyama (1982). In fact, Toomre-Hayashi model is a good approximation to a centrifugally supported disk formed by collapse of a rotating gas cloud (see, e.g., Nomura \&\ Mineshige 2000). Since the evolution of such systems may be driven by instabilities, Hanawa, Saigo, \&\ Matsumoto (2000) discussed the stability of the Toomre-Hayashi model against axisymmetric perturbations and Shu et al. (2000) analyzed the stability of this model under thin disk approximation while taking account of magnetic fields. The present analysis suggests that it is necessary to study the stability of our nonaxisymmetric disk solutions. Of course, the nonaxisymmetric solutions are nonmagnetized and the self-gravity balances with centrifugal force and gas pressure. But the magnetic field reduces the effective gravity and increases the sound speed (see, e.g., Li \&\ Shu 1997). The effects can be fully taken into account by replacing the gravitational constant $(G)$ and sound speed $(c_{s})$ with effective ones. So, the stability of magnetized disk is the same as that of nonmagnetized one. Also, one should perturb the self-similar equilibria and investigate their stability. The possibility of gravitational instability and its nonlinear consequence should be considered. Particularly, the nonaxisymmetric equilibria may fragment due to the gravitational instability and multiple stellar systems may form.

Since magnetic field, as well as rotation, plays an important role in star formation, an effect of purely toroidal magnetic field on the axisymmetric solutions has been studied. We assumed that the ratio of magnetic pressure to the thermal pressure, $\alpha$, is spatially constant, since we found this case to be integrable analytically. We showed that nonrotating and magnetized systems tend to the cylindrically symmetric configurations. On the other hand, MN showed that the nonrotating and nonmagnetized systems form ellipsoids (or spheres). If a system rotates such that $v_{0\varphi}<\sqrt{2\alpha}$, the system has a cylindrically symmetric structure. But if $v_{0\varphi}=\sqrt{2\alpha}$, the systems form ellipsoids or spheres. For $v_{0\varphi}>\sqrt{2\alpha}$ the axisymmetric equilibrium corresponds to a family of toroidal configurations. In fact, the magnetic field decreases the effect of rotation.

MN proved that the axisymmetric solutions with $B\neq 1$ are not force-free at two singular points, $r=0$ and $r=\infty$. In fact external forces hold the axisymmetric system in equilibrium. Are the nonaxisymmetric equilibria likely to be not force-free at singular points? Although we did not investigate this problem, it is likely that the nonaxisymmetric solutions are also not force-free at singular points. However, we think such configurations represent legitimate states of equilibrium. In this regard, we would like to remember the explanation of Galli et al. (2001): ``in the solutions, the (infinite) gravitational  force at the origin is exactly balanced by an (infinite) pressure gradient acting in concert  perhaps with an (infinite) centrifugal force. This balance is qualitatively no different than at any other point in the system , and it would be an artificial restriction to rule out eccentric equilibria simply because they have a nontrivial balance of forces at the origin rather than a trivial one.''

~\\

\noindent{\bf ACKNOWLEDGEMENTS}\\
We thank an anonymous referee for useful comments on the manuscript. We also wish to express our appreciation to Bruce Elmegreen for insightful comments and discussions. The research of MS is supported by Ferdowsi University. This work has made use of NASA's Astrophysical Data System Abstract Service.
~\\
~\\

\noindent{\bf REFERENCES}
\\Binney, J. \&\ Tremaine, S. 1987, Galactic Dynamics (Princeton: Princeton University Press)
\\Curry, C.L. 2000, ApJ, 541, 831
\\Fiege, J.D., \&\ Pudritz, R.E. 2000, MNRAS, 311, 85
\\Galli, D., Shu, F.H., Laughlin, G., \&\ Lizano, S. 2001, ApJ, in press
\\Hanawa, T., Saigo, K., \&\ Matsumoto, T. 2001, astro-ph/0010081
\\Hayashi, C., Narita, S., \&\ Miyama, S. M. 1982, Prog. Theor. Phys., 68, 1949
\\Heiles, C., Goodman, A.A., McKee, C.F., \&\ Zweibel, E.G. 1993, in Protostars and Planets III,ed. E. Levy \&\ J. Lunine (Tucson: University Arizona Press), 279
\\Jijina, J., Myers, P.C., \&\ Adams, F.C. 1999, ApJS, 125, 161
\\Kaifu, N. 1987, in {\it IAU Symposium 115, Star Forming Regions}, ed. M. Peimbert, and J. Jugaku (Dordrecht: Reidel), p. 275
\\Kiguchi, M., Narita, S., Miyama, S.M., \&\ Hayashi, C. 1987, ApJ, 317, 830
\\Kulkarni, S.R., \&\ Heiles, C. 1988, in Galactic and Extragalactic Radio Astronomy,ed. G.L. Verschuur \&\ K.I. Kellermann (Berlin: Springer-Verlag), 95
\\Li, Z.-Y., \&\ Shu, F.H. 1997, ApJ, 475, 237
\\Medvedev, M.V., \&\ Narayan, R. 2000, ApJ, 541, 579 (MN)
\\Nomura, H., \&\ Mineshige, S. 2000, ApJ, 536, 429
\\Ostriker, J. 1964, ApJ, 140, 1056
\\Schmid-Burgk, J. 1967, ApJ, 149, 727
\\Shu, F.H. 1992, Gas Dynamics (Mill Valley: University Science)
\\Shu, F.H., Laughlin, G., Lizano, S., \&\ Galli, D. 2000, ApJ, 535, 190
\\Stuart, J.T. 1967, J. Fluid Mech., 29, 417 
\\Tomisaka, K., Ikeuchi, S., \&\ Nakamura, T. 1988, ApJ, 326, 208
\\Toomre, A. 1982, ApJ, 259, 535 
\end{document}